\documentclass{article}

\usepackage{arxiv}

\usepackage[utf8]{inputenc} 
\usepackage[T1]{fontenc}    
\usepackage{hyperref}       
\usepackage{url}            
\usepackage{booktabs}       
\usepackage{amsfonts}       
\usepackage{nicefrac}       
\usepackage{microtype}      
\usepackage{lipsum}
\usepackage{multirow}
\usepackage{colortbl}
\usepackage{graphicx}
\usepackage[export]{adjustbox}
\usepackage[
  separate-uncertainty = true,
  multi-part-units = repeat
]{siunitx}

\title{A Stroke Detection and Discrimination Framework using Broadband Microwave Scattering on Stochastic Models with Deep Learning}

\author{
 Leeor Alon \\
  New York University School of Medicine\\
  Department of Radiology \\
  New York, NY 10016 \\
  \texttt{leeor.alon@nyumc.org} \\
  \And
    Seena Dehkharghani \\
  New York University School of Medicine\\
  Department of Radiology \\
  New York, NY 10016 \\
  \texttt{seena.dehkharghani@nyumc.org} \\
}

\begin{document}
\maketitle

\begin{abstract}
Stroke poses an immense public health burden and remains among the primary causes of death and disability worldwide. Emergent therapy is often precluded by late or indeterminate times of onset before initial clinical presentation. Rapid, mobile, safe and low-cost stroke detection technology remains a deeply unmet clinical need. Past studies have explored the use of microwave and other small form-factor strategies for rapid stroke detection; however, widespread clinical adoption remains unrealized. Here, we investigated the use of microwave scattering perturbations from ultra wide-band antenna arrays to learn dielectric signatures of disease. Two deep neural networks (DNNs) were used for: 1) stroke detection ("classification network"), and 2) characterization of the hemorrhage location and size ("discrimination network"). Dielectric signatures were learned on a simulated cohort of 666 hemorrhagic stroke and control subjects using 2D stochastic head models. The classification network yielded a stratified K-fold stroke detection accuracy of $\SI{94.6 \pm 2.86}{\percent}$ with an AUC of 0.996, while the discrimination network resulted in a mean squared error of <0.004 cm and <0.02 cm, for the stroke localization and size estimation, respectively. We report a novel approach to intelligent diagnostics using microwave wide-band scattering information thus circumventing conventional image-formation.
\end{abstract}


\section*{Introduction}
According to WHO estimates, stroke remains the second leading cause of death worldwide, and a leading cause for disability\cite{WHO2014}. CT and MRI are central to stroke care; however, safety concerns relating respectively to ionizing radiation and the use of strong magnetic fields, as well as limits to portability and high costs, preclude their practical use in most pre- and non-clinical settings\cite{Giroud2014worldwide,Mollura2014Radiology}. Further, profound disparities in the availability of such technologies perpetuate large imbalances in global health in the 21st century, disproportionately affecting underserved communities, with the WHO estimating nearly two-thirds of the world’s population as lacking access even to the most basic of medical imaging\cite{Ferns2011De}. The consequence of delayed access is astronomical, with 1.9 million neurons, 14 billion synapses, and twelve kilometers of myelinated white matter lost every minute, hence the axiom, \textit{time is brain}. Minimizing the time required for diagnosis and care for ischemic stroke is, therefore, paramount \cite{Saver2006Time}. Rupture of cerebral blood vessels, producing hemorrhagic stroke, may be similarly catastrophic and necessitates prompt diagnosis \cite{Hemphill2015}. In particular, subarachnoid hemorrhage, such as from rupture of cerebral aneurysms, is a potentially devastating injury occurring precipitously and often without warning in otherwise healthy individuals\cite{Lantigua2015Subarachnoid,Dority2016Subarachnoid}; including the sizeable percentage of those succumbing even before hospital arrival, 30-day mortality rates in some series exceed 40\% \cite{Chalouhi2013Review,Skodvin2017Cerebral}. Such diseases thus demand fast, accurate, and safe characterization to facilitate diagnosis, prognostication, and treatment selection \cite{Dority2016Subarachnoid,Lantigua2015Subarachnoid}.  

Magnetic Resonance Imaging (MRI) devices have been explored in acute-care settings \cite{Sheth2020}, but wide scale dissemination to the pre-clinical environment is limited by cost, technical and logistical constraints (e.g.shielding, size, safety, etc.), and access in underserved populations \cite{Geethanath2019}. Recently, several approaches for low-field (<0.3 T), low-cost, head MRI scanners have been proposed for acute-phase imaging (e.g. emergency departments, intensive care units, etc.). However, in the context of stroke detection, experience with such devices is preliminary and mobile deployment has yet to be fully realized \cite{Wald2020,Cooley2020}. Mobile Computerized Tomography (CT) units have been used in acute settings, but remain encumbered by concerns associated with ionizing radiation \cite{Miglioretti2017}, low sensitivity for hyper-acute ischemic stroke without the addition of iodinated contrasts, cost, and portability \cite{Fassbender2017}. 

Other imaging approaches have been explored for stroke detection, including measurement of electrical properties (EP) of tissues using transcranial currents; however, dielectric map reconstruction is limited by the localization of induced currents to the tissue surfaces and the ill-posedness of the reconstruction \cite{Brown2003}; consequently, deep-seated structures may not be unambiguously resolved with high fidelity, limiting anatomic and physiologic detail and clinical applicability \cite{Lin1982}. Mapping of EP alternatively with radio-frequency waves has emerged as a potential diagnostic and/or imaging approach, leveraging the propagation characteristics of electromagnetic waves in optically opaque media \cite{Lin1982}. Near-field microwave imaging (MI) systems estimate the spatial distribution of tissue EP by measuring the scattered field from an array of antennas, as initially described by Lin and Clark \cite{Lin1982} and subsequently by Semnov \cite{Semenov2005,Semenov2008}, Presson \cite{Trefna2008}, and Abbosh \cite{Mustafa2013}. 

MI methods can be divided into two broad categories: 1. quantitative imaging of the dielectric properties of tissues (conductivity and relative permeability) and 2. qualitative depiction of the geometrical features of the object as probed by the scattered fields \cite{Crocco2018}. Several potential solutions to the quantitative approach have been proposed according to physical models. These include the contrast source inversion (CSI) approach \cite{VanDenBerg1997}, experimentally revealing low resolution images for high dielectric contrasts \cite{Abubakar2002}, the linear inversion with truncated singular value decomposition (TSVD) and others \cite{Bertero1998}. However, the inherent non-linearity and ill-posedness of the inverse reconstruction problem remains challenging \cite{Crocco2018}, often necessitating regularization and/or parameter tuning. Finite Difference Time Domain (FDTD) electromagnetic field simulations have been used to ameliorate the MI reconstruction problem in which experimental data is compared against virtual simulations of the setup (solving Maxwell's equations \emph{in silico}). Since the inverse problem is ill-posed, due to the degrees of freedom exceeding the number of measured values, Tikhonov regularization \cite{Bertero1998} is applied. Modeling the architectural and dielectric complexity of the human brain present ongoing challenges \cite{Crocco2018}. By contrast, qualitative techniques, which aim to represent an image of the scattered field, have been proposed \cite{Colton1996,Kirsch1998,Catapano2007}, but lack information on the dielectric properties of tissues which may improve diagnostic value \cite{Crocco2018}.

Early work in MI hardware consisted of a vector network analyzer (VNA) connected to a computer control unit driving an antenna array. The forward and reflected power measured by each transmitting and receiving antenna is then used to calculate the S-parameter matrix, capturing the coupling between induced electromagnetic field and the head. Often a matching medium is used to improve energy transfer from the antennas to the head. Initial works utilized narrow-band directional antennas \cite{Lin1982} or open wave-guides \cite{Semenov2008}, often operating between 0.8 and 1.2 GHz. These frequencies were chosen due to the electromagnetic wave penetration and tissue dielectric contrast at these frequencies, enabling probing of changes in dielectric properties in deep-seated regions with ample spatial resolution \cite{Crocco2018}. More recently, wide-band systems (typically under 6 GHz) have become more mainstream, providing greater information about the electrical disturbances and obtained with larger frequency sweeps \cite{Jamlos2015,Fei2016,Alqadami2019}. Software defined radio (SDR) technology has further advanced MI by reducing the cost of signal transmission and reception technology \cite{Stancombe2019}. 

The ongoing need for low cost, rapidly deployable, and scalable solutions for preclinical stroke diagnosis motivates the development of novel stroke detection methods. Here, we present a framework for hemorrhagic stroke detection using learned scattering signatures from ultra wide-band antenna arrays between 0.5 and 6 GHz. While it is well established that changes in the dielectric properties of the brain are altered with disease, we investigated whether microwave scattering perturbations can be used to learn disease \emph{signatures}. The potential to learn the dielectric signatures of strokes using big-data, on a population level, remains, to the best of our knowledge, unrealized. Herein, deep neural networks (DNNs) in combination with scattering information from microwave sources were used to classify and discriminate the presence of stroke. Results were shown for stochastic \emph{in silico}, multi-tissue head models representing the  dielectric disturbances occurring in the setting of hemorrhagic stroke and derived from electromagnetic field simulations. The results introduce a novel approach for intelligent diagnostics, circumventing conventional image-formation.

\section*{Methods}
\label{sec:methods}
\subsection*{Electromagnetic (EM) field simulations}
\begin{figure}[t]
  \includegraphics[width=0.8\textwidth, center]{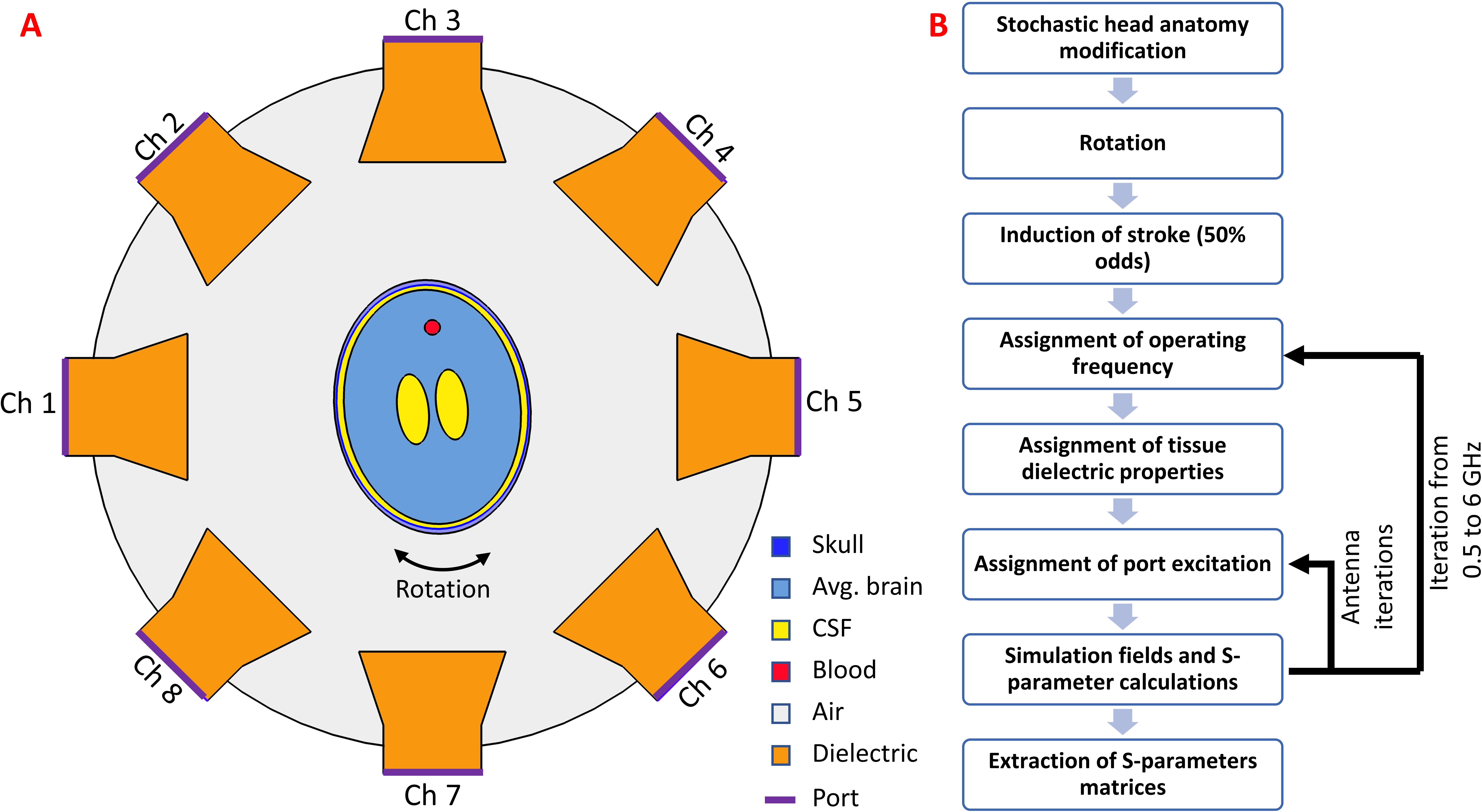}
  \caption{A. Simulation setup for for FEM solver. 8-horn antennas were placed around varying 2D multi-tissue head models B. Flow chart for processing of EM field simulations, allowing modification of the head models, dielectric properties, and induction of stroke, simulating different frequencies and excitations.}
  \label{fig:fig1}
  \centering
\end{figure}

Simulations where conducted using the COMSOL finite element method (FEM) software (COMSOL Multiphysics, Burlington, MA USA) on an array of 8 wide-band horn antennas, each with an aperture of 120 mm and driven using a waveguide excitation. Each antenna was placed azimuthally around head models at 45 degree increments and filled with a dielectric with relative permittivity of 50 to reduce the cutoff frequency. The setup is summarized in figure \ref{fig:fig1}A. A two-dimensional (2D) solver was used to reduce computational time as a large number of simulations were necessary for the varying anatomy-coil-frequency conditions. Simulations were conducted by alternating with a single active port while recording the voltage on every port to calculate the S-parameter matrix at each frequency. A stationary frequency domain direct solver using the Newton method was used to ensure convergence. A total of 20 frequency sweeps were conducted at equal increments between 0.5 and 6 GHz and were prescribed to inform DNNs trained for classification and discrimination of stroke. After the simulations concluded, random Gaussian noise with mean zero and standard deviation of -100 dBm was added to the S-parameters; this noise was introduced to represent the noise floor in commonly-used network analyzers \cite{Aglinet2000} such as can be incorporated in a system design to acquire S-parameters for stroke detection \cite{Mustafa2013}. For each head simulation a complex matrix with dimension equal to: number of channels x number of channels x number of frequencies was obtained. EM field computations and post-processing were executed on a simulation workstation with a dual-core Xeon E5-2630 CPU with 128 GB of RAM. 

\subsection*{Head EP Perturbation}
A total of 666 head models were generated in simulation using COMSOL Livelink, enabling scripting within the Matlab (Mathworks, Natick, Massachusetts USA) environment. A 2D synthetic, trans-axial slice through the brain, was modelled based on the 3D Specific Anthropomorphic Mannequin (SAM) model \cite{GORDON1988}, producing a simulated central slice through the brain at the level of the lateral ventricles. Stochastic modifications to the brain dimensions were applied to each head model, comprising variation in the width and length of the head and brain independently, and width and length of the lateral ventricles. The head models were placed centrally within the antenna array, with random reorientation of the head prescribed relative to the array, in order to represent varying head orientations occurring when the head is fixated to a plastic stereotactic frame permitted to pivot relative to the antenna array, yet remains immobile with respect to translation. Variations in head anatomy and rotation are summarized in table \ref{table:tbl1}.
A total of five different tissues were included in the simulations: skull, blood, cerebrospinal fluid (CSF), average brain, and air. For each frequency of interest between 0.5 and 6 GHz, the dielectric properties were extracted from Gabriel et al. \cite{Gabriel1996,Gabriel1996a,Gabriel1996b}. Random-valued head anatomies were fed to the DNN representing subject-wise variation of the head, thus presenting the DNN with highly varying conditions that disrupt the S-parameter matrices to facilitate learning. Fifty percent of the heads were selected at random and intracranial blood was introduced in random locations, simulating intraparenchymal hemorrhagic strokes. Sizes of the strokes were independently varied in the x- and y-directions with a 2$\pm$0.5cm (mean$\pm$standard deviation) in diameter (ranging between 0.5 and 4 cm, in the x- and y-directions, independently) (table \ref{table:tbl1}). The stroke x- and y-locations were selected from a uniform random distribution, with the outer bounds of the distribution changing depending on the size of the brain. A total of 106,560 simulations were conducted using different head anatomies, frequency sweeps, and antennas driven. Execution of these simulations was performed in >720 continuous hours of simulation time. The pipeline for stochastic head model simulations and computation of the S-parameter matrices at each frequency of operation is summarized in figure \ref{fig:fig1}B. The effect of perturbations on the dielectric properties of the head induced by different strokes was analyzed and plotted across different frequencies. 

\begin{figure}
  \includegraphics[width=1\textwidth, center]{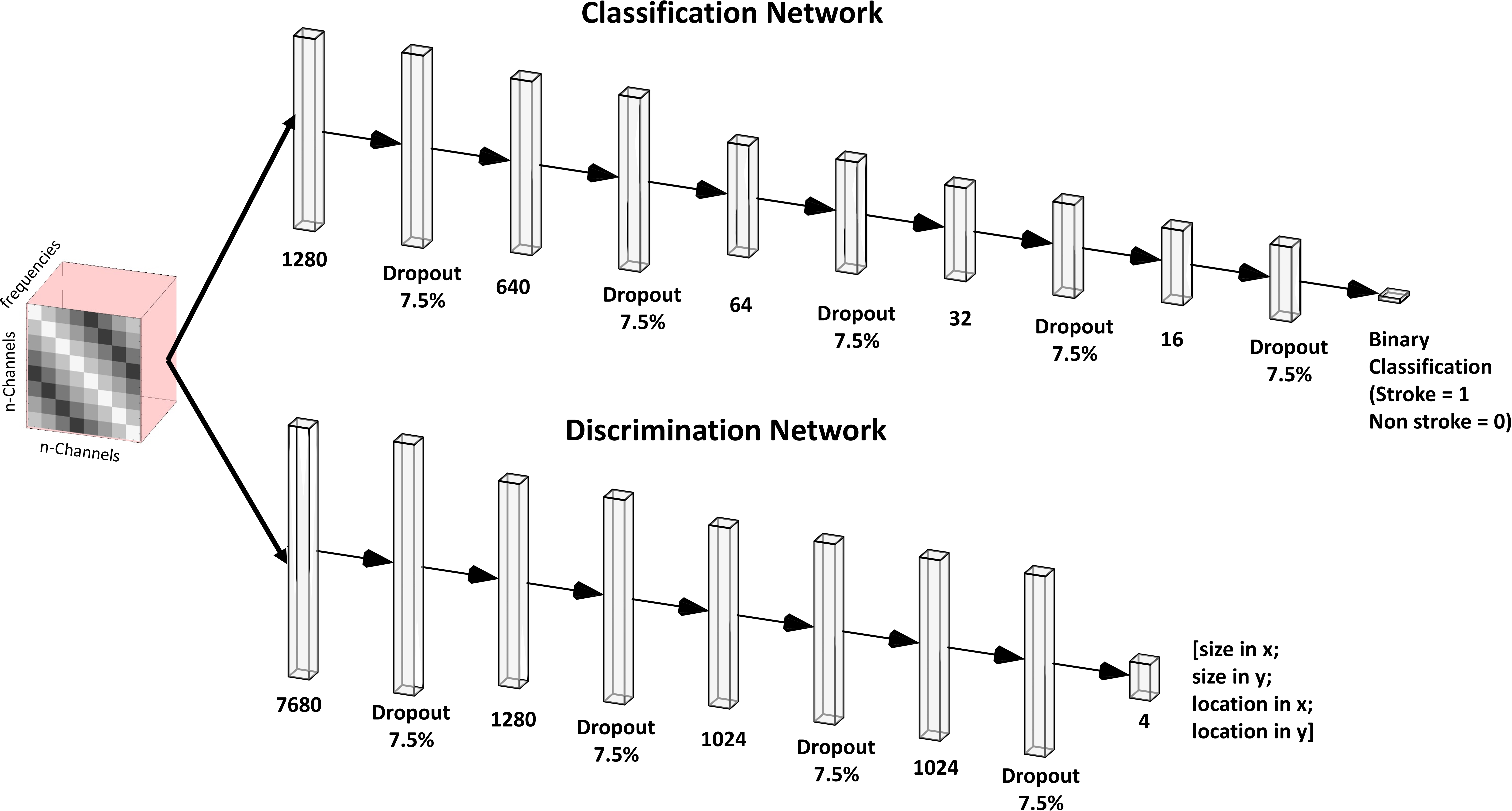}
  \caption{DNN architectures used for the classification and discrimination networks. Amplitude and phase of the multi-frequency S-parameter matrices were concatenated, vectorized and fed into each DNN. A sigmoid output and 4-vector linear output were used for the classification and discrimination networks, respectively.}
  \label{fig:fig2}
\end{figure}

\begin{table}[b]
\centering
\caption{Head anatomy variation}
\arrayrulecolor{black}
\begin{tabular}{|c|l|l|l|l|l|l|}
\hline
\textbf{Anatomy} & \multicolumn{1}{c|}{\textbf{Orientation}} & \multicolumn{1}{c|}{\textbf{Mean}} & \multicolumn{1}{c|}{\textbf{Std.}} & \textbf{Min} & \textbf{Max} & \multicolumn{1}{c|}{\textbf{Unit}} \\ \hline
\multirow{2}{*}{\textbf{Head size}} & x-direction & 20 & 2 & 13.8 & 25.6 & cm \\ \cline{2-7} 
 & y-direction & 26 & 2.6 & 19.8 & 33.0 & cm \\ \hline
\multirow{2}{*}{\textbf{Vent size}} & x-direction & 4.55 & 0.5 & 5.9 & 3.5 & cm \\ \cline{2-7} 
 & y-direction & 8.45 & 0.85 & 10.8 & 6.34 & cm \\ \hline
\multicolumn{1}{|l|}{\textbf{Head rot.}} & xy-plane & 0 & 3.5 & -10.76 & 10.18 & deg. \\ \hline
\multirow{2}{*}{\textbf{Blood size}} & x-direction & 2 & 0.5 & 0.84 & 3.8 & cm \\ \cline{2-7} 
 & y-direction & 2 & 0.5 & 0.84 & 3.8 & cm \\ \hline
\end{tabular}
\label{table:tbl1}
\end{table}

\subsection*{Classification and Discrimination}
Multi-frequency S-parameter matrices were exported into python. The dimensions of each complex-valued matrix were n\textsubscript{ch} x n\textsubscript{ch} x n\textsubscript{f}, where n\textsubscript{ch} is the number of channels and n\textsubscript{f} is the number of simulated frequencies. Amplitude and phase of the complex-valued matrices were reformatted, for each model, to n\textsubscript{ch}\textsuperscript{2} * n\textsubscript{f}. Training, testing, and validation of the neural net models were conducted in the Keras environment \cite{Chollet2015}. Before training, the data across subjects were split such that 70\% of the data were used for training, 15\% for validation, and 15\% for testing, respectively. Training, testing, and validation data sets were normalized using a standard scalar normalizer such that the data was scaled to unit variance. Two distinct DNNs were used for the stroke diagnosis: the first DNN was used to classify the presence of hemorrhagic stroke using S-parameter matrix data ("Classification DNN") and the second DNN was used to discriminate the stroke location and size inside the head ("Discrimination DNN"). The classification DNN was trained using 666 stroke and control head models, consisting of 307 simulated strokes (46\%). The discrimination DNN was trained using the stroke-only cohort. A graphical depiction of the classification DNN comprising fully-connected layers followed by a final sigmoid activation layer is shown in figure \ref{fig:fig2}, top. Input to the classification network comprised 1280 unique parameters (generated from the complex s-parameter matrices) and an input layer for the classification network accordingly set with the purpose of identifying frequency dependent coil-coupling relationships that are unique for identification of stroke. The discrimination DNN comprising a wider fully-connected architecture terminating in a 4-dimensional linear layer with the 4-dimensional vector representing the x-, y-locations and size of hemorrhage in the x- and y-directions, respectively, thus characterizing the dimension and location of the hemorrhage \ref{fig:fig2}, bottom. For both the classification and discrimination networks a dropout layer with dropout threshold of 7.5\% was introduced between the fully connected layers to reduce over-fitting. Each layer of the classification network was initialized using a truncated normal distribution with zero mean and standard deviation of 0.75 and the discrimination network was initialized using a truncated normal distribution with zero mean and standard deviation of 0.05. Model optimization for both networks was performed using the Adam optimizer \cite{Kingma2015}. A binary cross-entropy loss function \cite{IanGoodfellowYoshuaBengio2015} with 20 epochs and a batch size of 100 was used for the classification network training, while a mean square error (MSE) loss function, 1000 epochs, and a batch size of 100 were used for discrimination network training. The loss function and accuracy of each trained network as a function of epoch was plotted. Classification accuracy was quantified using the stratified K-fold method\cite{Salzberg1997} with 10 splits. A Receiver-Operating Characteristic (ROC) curve was plotted and the Area Under the Curve (AUC) was computed. The discrimination network was evaluated by calculation of the test-set MSE for the errors in predicted x-, y- locations and x-,y- stroke size. For visualization purposes, results were compared against the ground truth in randomly-chosen stroke models selected from the test-set. Frequency- and antenna- dependent stroke dielectric effects were assessed by plotting changes in the magnitude of the S-parameter matrices occurring due to the presence of stroke for 0.5, 1 and 3 GHz and presented dB scale.

\section*{Results}
\label{sec:results}

\subsection*{Classification}
\begin{figure}
  \includegraphics[width=1\textwidth, center]{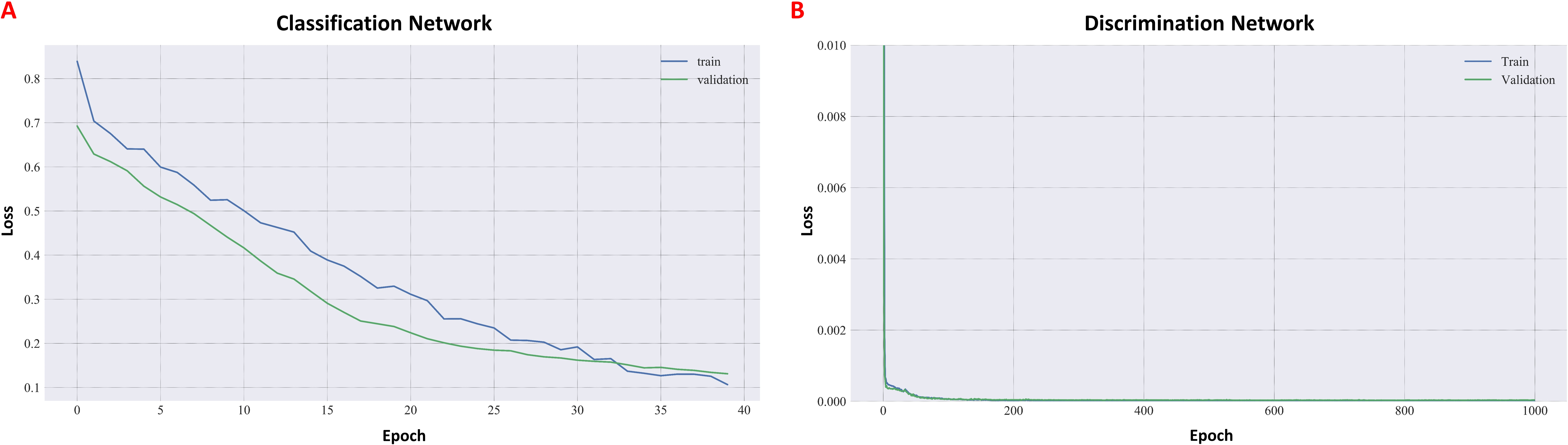}
  \caption{Classification network loss as a function of epoch number for the training and validation sets (A) and Training Discrimination network MSE as a function of epoch number for the stroke-only train and validation sets.}
  \label{fig:fig3}
\end{figure}

\begin{figure}[b]
  \includegraphics[width=1\textwidth, center]{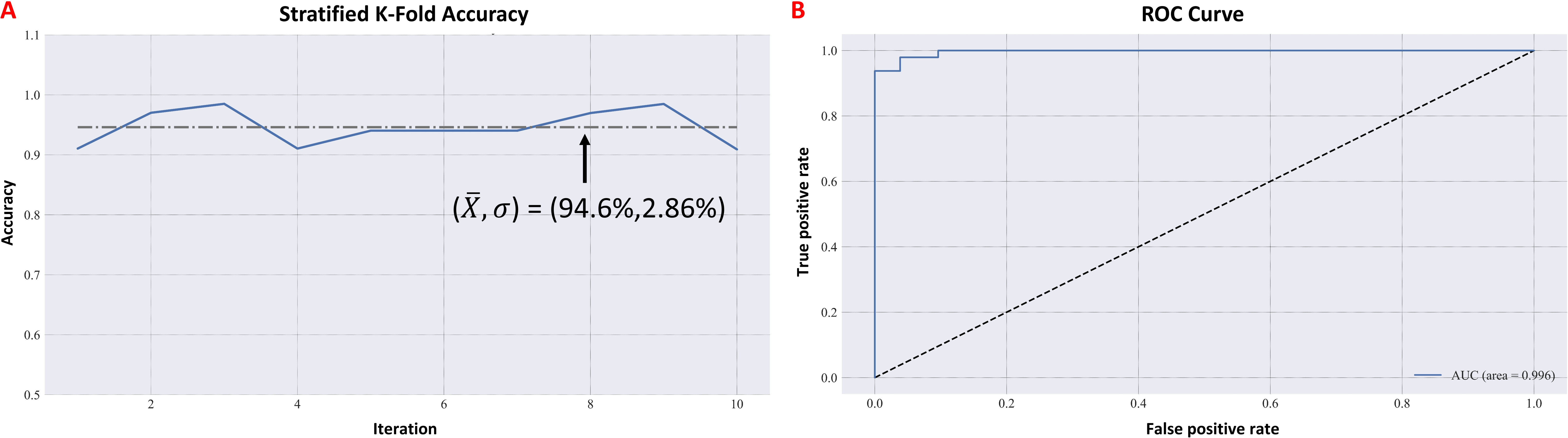}
  \caption{Stratified K-fold accuracy as a function of iteration for the classification network (blue); mean and standard deviation shown in black (A) and ROC curve computed using the test set predicted stroke classifications (B).}
  \label{fig:fig4}
\end{figure}

Execution of the classification network training was performed and the binary cross-entropy loss as a function of epoch was plotted as shown in figure \ref{fig:fig3}A, demonstrating convergence of the network. Generalization was evaluated using the stratified K-fold method\cite{Salzberg1997}, exhibiting a mean model accuracy of 94.6\% with a standard deviation of $\pm$2.86\% (figure \ref{fig:fig4}A). Test set ROC curve is shown in figure \ref{fig:fig4}B, demonstrating an AUC of 0.996 for the detection of stroke, suggesting that the classification DNN generalizes well using broadband S-parameter measurements as inputs. 

\subsection*{Discrimination}
\begin{figure}[t]
  \includegraphics[width=0.8\textwidth, center]{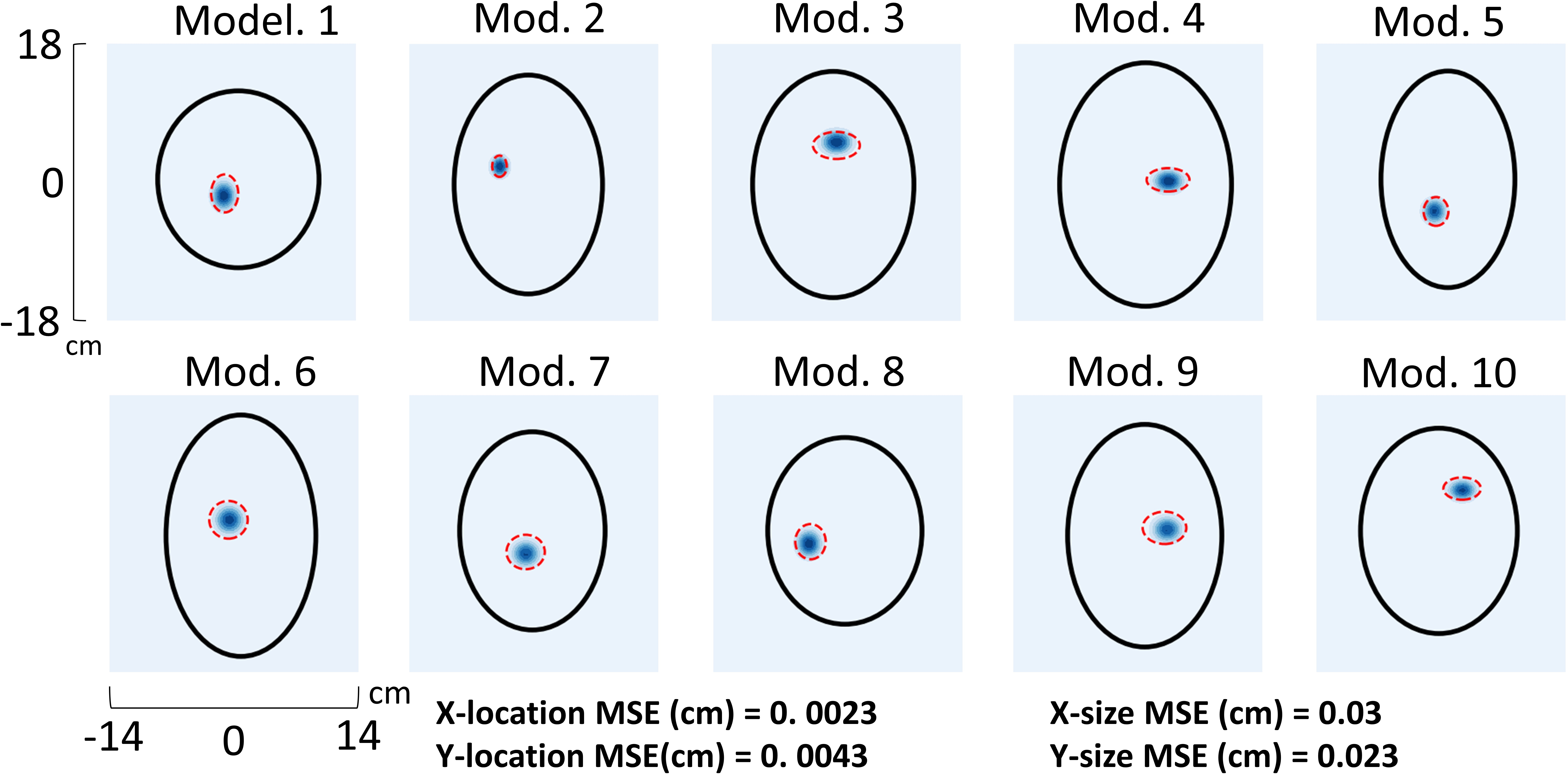}
  \caption{Test set stroke predicted by the discrimination network. True stroke size and location(red dotted line) is compared against the predicted strokes (blue Gaussians)}
  \label{fig:fig5}
\end{figure}

The discrimination network was designed to predict both the size and location of the stroke; discrimination was performed via learning patterns from frequency dependent S-parameters from different antenna elements, rather than relying on conventional microwave imaging reconstruction. In order to better discriminate stroke spatial characteristics, a larger network with more redundancy was required, consisting of a 6x larger first layer, compared to the classification network, and a relatively larger number of neurons in the subsequent hidden layers, permitting finer prediction of size and location (figure \ref{fig:fig2}, bottom). The discrimination network MSE as a function of epoch is shown in figure \ref{fig:fig3}B, demonstrating close agreement between the test and validation set errors. Upon convergence of training, size and location of the strokes in 10 random head models from the test set cohort were plotted in figure \ref{fig:fig5}, where the dotted red ovals represent the in silico stroke. As shown, varying head sizes as well as stroke locations and sizes were were used as inputs to the DNN. For clarity of representation, additional anatomic features (other than the stroke) were excluded. The predicted stroke size and location predicted by the DNN is illustrated as a 2D Gaussian probability density function shown in blue. Test set MSE for stroke localization in the x- and y-directions was 0.0023 cm and 0.0043 cm, respectively, and stroke size MSE in the x- and y- directions was 0.03 cm and 0.023 cm, respectively.

\subsection*{Frequency and antenna dependent dielectric effects}
The observed changes in the S-parameters for three different strokes (represented as red ovals A, B and C) are summarized in figure \ref{fig:fig6} for different frequencies. 
Lesion A exhibited prominent changes in the S-parameters adjacent to antennas 4 and 5 occurring above 0.5 GHz. The change in S45 represents the increase in power transfer between antennas 4 and 5 occurring due to the changes in dielectric parameters in the brain induced by the presence of stroke. When the same-sized lesion (B), is displaced to an deeper region of the brain, changes in the S-parameters of channels 4, 5 and 6 are observed at the 0.5 GHz, as well as at higher frequencies. A larger lesion C, in the contralateral side (relative to strokes A and B), changed the S-parameter values for channels 1 and 2 (closest to lesion C) and their respective coupling.

\section*{Discussion}
\label{sec:discussion}
\begin{figure}[t]
  \includegraphics[width=1\textwidth, center]{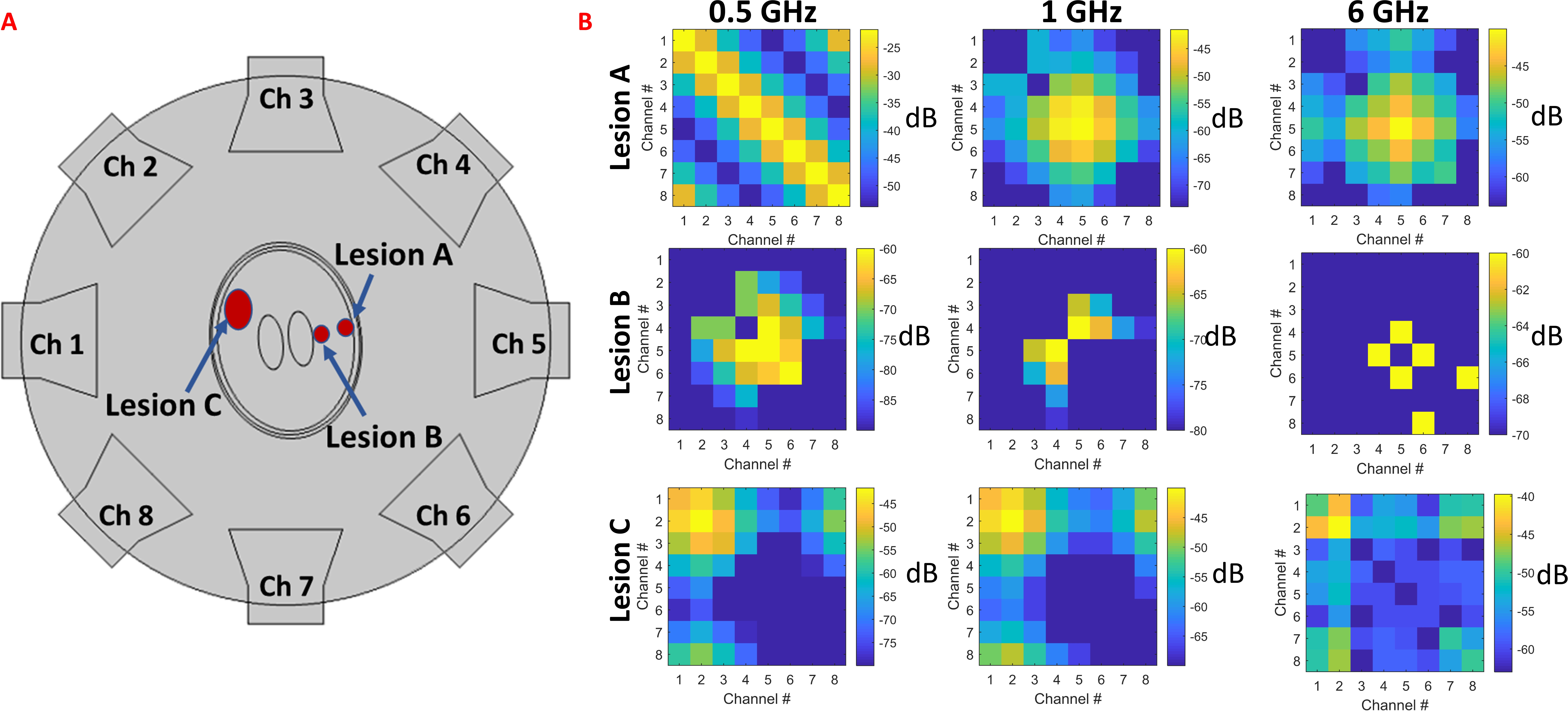}
  \caption{Frequency- and space- dependent changes in S-parameter values for different lesions}
  \label{fig:fig6}
\end{figure}

In this work we demonstrated that physical changes in tissue EPs caused by hemorrhagic stroke cause measurable changes in EM wave scattering. These changes in the S-parameter matrices can be used to train networks for intelligent diagnosis and discrimination of stroke. Past studies have utilized narrow-band frequency excitation and reception systems \cite{Semenov2005,Abubakar1997,Mojabi2012} motivated by the reported advantages of dielectric contrast in the brain at those frequencies; here, broadband acquisitions between 0.5 and 6 GHz provided complementary information. Specifically, the importance of multi-antenna and multi-frequency information as was depicted in figure \ref{fig:fig6} underscored that higher S-parameter disruptions at higher frequencies occur due to the higher sensitivity in shallow depths, due to the skin effect, as well as the size of the lesion. These changes provided valuable information for the DNNs to facilitate learning the presence of stroke as well as the capability to localize and estimate the size of the stroke. In this respect, this study presents a first-of-its-kind approach whereby the additional contrast arising from a broadband frequency sweep in a simulated cohort could be utilized to facilitate learning and generalization, while circumventing conventional MI reconstruction inverse problems. 

Two-dimensional EM simulations were employed due to computational demands inherent to three-dimensional simulations. Here, more than 100,000 simulations were conducted in order to calculate electromagnetic fields and thus S-parameters across all frequency, subject, and antenna permutations, requiring approximately one month of continuous computation. The challenges inherent to engineering realistic, 3D patho-anatomic models for electromagnetic simulation are well-recognized and constitute ongoing motivation for the development of robust experimental methods. Reduction of the dimensionality to two-dimensions to alleviate computational demand consequently eliminated the z-dimension (i.e. the craniocaudal direction) thus converting the hemorrhage to an affected 2D area of the simulated brain with Maxwell's equations solved in 2D. Notwithstanding, we anticipate that a transition to three dimensions in realizable antenna array systems would benefit from the utilization of a greater number of antennas, thus capturing 3D spatial detail such as that arising from a 3D hemorrhage volume within the head. While past studies have demonstrated that large head-encircling arrays (>30 antennas) together with electromagnetic field simulations can be used for stroke detection \cite{Semenov2009,Semenov2017}, the benefits of machine learning approaches applied to large cohorts, including with broadband frequency interrogation, remain untested. 

In this study we explore new avenues for machine learning-assisted MI, leveraging the additional contrast arising from broadband microwave frequencies, in order to improve prediction of the presence, location, and size of intracranial hemorrhage. Importantly, however, the optimal antenna configuration and prescription of interrogated frequencies requires further study to balance sensitivity with hardware and computational demand. We anticipate that tolerance to reductions will most likely be hardware- and application-specific and a topic for future study. The reduction in the number of frequency sweeps and/or coils used will result in acceleration of S-parameter acquisitions to improve patient comfort or time to diagnosis. Further, our selection of a simulated cohort of 666 stroke and control subjects constitutes only an initial approximation of an achievable cohort in contemporary stroke trials and is representative of stroke imaging workflow in our own practice.

Acquisition systems capable of realizing this approach could potentially employ wide-band antenna arrays encircling the patient's head. Each antenna can be connected to controllable RF/microwave switches routing each antenna to receive or transmit signals and connected to a multi-port vector network analyzer (VNA). Alternatively, SDRs can be utilized alongside directional couplers to gather wide-band scattering information. We expect that such systems will be two orders of magnitude less expensive than state of the art imaging systems (e.g. clinical MRI).

We elected to simulate intracerebral hemorrhage in this initial investigation. Blood provides a stronger dielectric contrast than ischemia, which may be more pathologically variegated at the tissue level, challenging development of realistic simulated conditions. While this study did not attempt to classify or discriminate ischemic stroke, the detection accuracy for hemorrhage >1cm in diameter (see table \ref{table:tbl1} for stroke variation in x- and y- directions) was 94.6\%, and therefore the dependable exclusion of hemorrhage can serve as a valuable screening tool in the acute stroke setting to facilitate rapid thrombolysis as per current guidelines\cite{Powers2019} in early-presenting patients. This latter point bears emphasis, as intravenous thrombolysis remains the only approved pharmacologic therapy in acute ischemic stroke patients, but with rare exceptions cannot be administered beyond the initial, hyperacute window. Consequently, only ~12\% of patients receive critical intravenous thrombolysis, despite a 30\% improvement in outcomes among those treated. Future studies will, nevertheless, focus on the detection of ischemic stroke as well as other neurologic disorders using the \emph{dielectrographic} technique described here.

Overall, we present a method for hemorrhagic stroke detection and discrimination by means of learning disease dielectric signatures from S-parameter measurements using wide-band antenna arrays. We investigated whether microwave scattering perturbations can be used to learn disease signatures from a large simulated cohort. Two DNN architectures were designed and tuned for classification and discrimination in presence of stroke in stochastic \emph{in silico}, multi-compartment head models representing the tissue dielectric disturbances occurring in the setting of hemorrhagic stroke and derived from electromagnetic field simulations. The nature of our proposed approach provides an effective potential strategy for rapid stroke detection and management, circumventing conventional image generation.

\newpage

\bibliographystyle{unsrt}  

\bibliography{references}  

\begin{thebibliography}{10}

\bibitem{WHO2014}
World~Health Organisation.
\newblock Global status report on noncommunicable diseases 2014.
\newblock Technical report, World Health Organisation, 2014.

\bibitem{Giroud2014worldwide}
Maurice Giroud, Agnès Jacquin, and Yannick Béjot.
\newblock The worldwide landscape of stroke in the 21st century.
\newblock {\em The Lancet}, 2014.

\bibitem{Mollura2014Radiology}
Daniel~J. Mollura and Matthew~P Lungren.
\newblock {\em Radiology in Global Health: Strategies, Implementation, and
  Applications}.
\newblock Springer, 2014.

\bibitem{Ferns2011De}
Sandra~P. Ferns, Marieke~E.S. Sprengers, Willem Jan~J. Van~Rooij, René Van
  Den~Berg, Birgitta~K. Velthuis, Gérard~A.P. De~Kort, Menno Sluzewski, Wim~H.
  Van~Zwam, Gabriël~J.E. Rinkel, and Charles~B.L.M. Majoie.
\newblock De novo aneurysm formation and growth of untreated aneurysms: A
  5-year mra follow-up in a large cohort of patients with coiled aneurysms and
  review of the literature.
\newblock {\em Stroke}, 2011.

\bibitem{Saver2006Time}
Jeffrey~L. Saver.
\newblock Time is brain - quantified.
\newblock {\em Stroke}, 2006.

\bibitem{Hemphill2015}
J.~Claude Hemphill, Steven~M. Greenberg, Craig~S. Anderson, Kyra Becker,
  Bernard~R. Bendok, Mary Cushman, Gordon~L. Fung, Joshua~N. Goldstein, R.~Loch
  Macdonald, Pamela~H. Mitchell, Phillip~A. Scott, Magdy~H. Selim, and Daniel
  Woo.
\newblock {Guidelines for the Management of Spontaneous Intracerebral
  Hemorrhage}.
\newblock {\em Stroke}, 2015.

\bibitem{Lantigua2015Subarachnoid}
Hector Lantigua, Santiago Ortega-Gutierrez, J.~Michael Schmidt, Kiwon Lee,
  Neeraj Badjatia, Sachin Agarwal, Jan Claassen, E.~Sander Connolly, and
  Stephan~A. Mayer.
\newblock Subarachnoid hemorrhage: Who dies, and why?
\newblock {\em Critical Care}, 2015.

\bibitem{Dority2016Subarachnoid}
Jeremy~S. Dority and Jeffrey~S. Oldham.
\newblock Subarachnoid hemorrhage: An update.
\newblock {\em Anesthesiology Clinics}, 2016.

\bibitem{Chalouhi2013Review}
Nohra Chalouhi, Brian~L. Hoh, and David Hasan.
\newblock Review of cerebral aneurysm formation, growth, and rupture.
\newblock {\em Stroke}, 2013.

\bibitem{Skodvin2017Cerebral}
Torbjørn~Øygard Skodvin, Liv-Hege Johnsen, Øivind Gjertsen, Jørgen~Gjernes
  Isaksen, and Angelika Sorteberg.
\newblock Cerebral aneurysm morphology before and after rupture.
\newblock {\em Stroke}, 2017.

\bibitem{Sheth2020}
Kevin~N. Sheth, Mercy~H. Mazurek, Matthew~M. Yuen, Bradley~A. Cahn, Jill~T.
  Shah, Adrienne Ward, Jennifer~A. Kim, Emily~J. Gilmore, Guido~J. Falcone,
  Nils Petersen, Kevin~T. Gobeske, Firas Kaddouh, David~Y. Hwang, Joseph
  Schindler, Lauren Sansing, Charles Matouk, Jonathan Rothberg, Gordon Sze,
  Jonathan Siner, Matthew~S. Rosen, Serena Spudich, and W.~Taylor Kimberly.
\newblock {Assessment of brain injury using portable, low-field magnetic
  resonance imaging at the bedside of critically ill patients}.
\newblock {\em JAMA Neurology}, 02114(1):E1--E7, 2020.

\bibitem{Geethanath2019}
Sairam Geethanath and John~Thomas Vaughan.
\newblock {Accessible magnetic resonance imaging: A review}.
\newblock {\em Journal of Magnetic Resonance Imaging}, 49(7):e65--e77, 2019.

\bibitem{Wald2020}
Lawrence~L. Wald, Patrick~C. McDaniel, Thomas Witzel, Jason~P. Stockmann, and
  Clarissa~Zimmerman Cooley.
\newblock {Low-cost and portable MRI}.
\newblock {\em Journal of Magnetic Resonance Imaging}, 52(3):686--696, 2020.

\bibitem{Cooley2020}
Clarissa~Z. Cooley, Patrick~C. McDaniel, Jason~P. Stockmann, Sai~Abitha
  Srinivas, Stephen~F. Cauley, Monika {\'{S}}liwiak, Charlotte~R. Sappo,
  Christopher~F. Vaughn, Bastien Guerin, Matthew~S. Rosen, Michael~H. Lev, and
  Lawrence~L. Wald.
\newblock {A portable scanner for magnetic resonance imaging of the brain}.
\newblock {\em Nature Biomedical Engineering}, 2020.

\bibitem{Miglioretti2017}
Diana~L Miglioretti, Eric Johnson, Andrew Williams, T~Robert, Sheila Weinmann,
  Leif~I Solberg, Heather Spencer, Douglas Roblin, Michael~J Flynn, and
  Nicholas Vanneman.
\newblock {Pediatric Computed Tomography and Associated Radiation Exposure and
  Estimated Cancer Risk}.
\newblock {\em Jama perdiactrics}, 167(8):700--707, 2013.

\bibitem{Fassbender2017}
Klaus Fassbender, James~C. Grotta, Silke Walter, Iris~Q. Grunwald, Andreas
  Ragoschke-Schumm, and Jeffrey~L. Saver.
\newblock {Mobile stroke units for prehospital thrombolysis, triage, and
  beyond: benefits and challenges}.
\newblock {\em The Lancet Neurology}, 16(3):227--237, 2017.

\bibitem{Brown2003}
B.~H. Brown.
\newblock {Electrical impedance tomography (EIT): A review}, 2003.

\bibitem{Lin1982}
James~C. Lin and Martin~J. Clarke.
\newblock {Microwave Imaging of Cerebral Edema}.
\newblock {\em Proceedings of the IEEE}, 70(5):523--524, 1982.

\bibitem{Semenov2005}
Serguei~Y. Semenov, Alexander~E. Bulyshev, Aria Abubakar, Vitally~G. Posukh,
  Yuri~E. Sizov, Alexander~E. Souvorov, Peter~M. {Van Den Berg}, and Thomas~C.
  Williams.
\newblock {Microwave-tomographic imaging of the high dielectric-contrast
  objects using different image-reconstruction approaches}.
\newblock {\em IEEE Transactions on Microwave Theory and Techniques}, 2005.

\bibitem{Semenov2008}
Serguei~Y. Semenov and Douglas~R. Corfield.
\newblock {Microwave Tomography for Brain Imaging: Feasibility Assessment for
  Stroke Detection}.
\newblock {\em International Journal of Antennas and Propagation}, 2008.

\bibitem{Trefna2008}
Hana Trefn{\'{a}} and Mikael Persson.
\newblock {Antenna array design for brain monitoring}.
\newblock In {\em 2008 IEEE International Symposium on Antennas and Propagation
  and USNC/URSI National Radio Science Meeting, APSURSI}, 2008.

\bibitem{Mustafa2013}
S.~Mustafa, B.~Mohammed, and A.~Abbosh.
\newblock {Novel preprocessing techniques for accurate microwave imaging of
  human brain}.
\newblock {\em IEEE Antennas and Wireless Propagation Letters}, 2013.

\bibitem{Crocco2018}
Lorenzo Crocco, Irene Karanasiou, Michael~L. James, and Raquel~Cruz
  Conceição.
\newblock {\em {Emerging electromagnetic technologies for brain diseases
  diagnostics, monitoring and therapy}}.
\newblock Springer, 2018.

\bibitem{VanDenBerg1997}
Peter~M. {Van Den Berg} and Ralph~E. Kleinman.
\newblock {A contrast source inversion method}.
\newblock {\em Inverse Problems}, 13(6):1607--1620, 1997.

\bibitem{Abubakar2002}
Aria Abubakar, Peter~M. {Van Den Berg}, and Jordi~J. Mallorqui.
\newblock {Imaging of biomedical data using a multiplicative regularized
  contrast source inversion method}.
\newblock {\em IEEE Transactions on Microwave Theory and Techniques},
  50(7):1761--1771, 2002.

\bibitem{Bertero1998}
Mario Bertero and Patrizia Boccacci.
\newblock {\em {Introduction to Inverse Problems in Imaging}}.
\newblock CRC Press, 1998.

\bibitem{Colton1996}
David Colton and Andreas Kirsch.
\newblock {A simple method for solving inverse scattering problems in the
  resonance region}.
\newblock {\em Inverse Problems}, 12(4):383--393, 1996.

\bibitem{Kirsch1998}
Andreas Kirsch.
\newblock {Characterization of the shape of a scattering obstacle using the
  spectral data of the far field operator}.
\newblock {\em Inverse Problems}, 14(6):1489--1512, 1998.

\bibitem{Catapano2007}
Ilaria Catapano, Lorenzo Crocco, and Tommaso Isernia.
\newblock {On simple methods for shape reconstruction of unknown scatterers}.
\newblock {\em IEEE Transactions on Antennas and Propagation},
  55(5):1431--1436, 2007.

\bibitem{Jamlos2015}
M.~A. Jamlos, M.~F. Jamlos, and A.~H. Ismail.
\newblock {High performance novel UWB array antenna for brain tumor detection
  via scattering parameters in microwave imaging simulation system}.
\newblock {\em 2015 9th European Conference on Antennas and Propagation, EuCAP
  2015}, pages 3--7, 2015.

\bibitem{Fei2016}
Chunlong Fei, Chi~Tat Chiu, Xiaoyang Chen, Zeyu Chen, Jianguo Ma, Benpeng Zhu,
  K.~Kirk Shung, and Qifa Zhou.
\newblock {Ultrahigh frequency (100 MHz-300 MHz) ultrasonic transducers for
  optical resolution medical imagining}.
\newblock {\em Scientific Reports}, 6(June):1--8, 2016.

\bibitem{Alqadami2019}
Abdulrahman~S.M. Alqadami, Konstanty~S. Bialkowski, Ahmed~Toaha Mobashsher, and
  Amin~M. Abbosh.
\newblock {Wearable Electromagnetic Head Imaging System Using Flexible Wideband
  Antenna Array Based on Polymer Technology for Brain Stroke Diagnosis}.
\newblock {\em IEEE Transactions on Biomedical Circuits and Systems},
  13(1):124--134, 2019.

\bibitem{Stancombe2019}
Anthony~Edgar Stancombe, Konstanty~S. Bialkowski, and Amin~M. Abbosh.
\newblock {Portable microwave head imaging system using software-defined radio
  and switching network}.
\newblock {\em IEEE Journal of Electromagnetics, RF and Microwaves in Medicine
  and Biology}, 3(4):284--291, 2019.

\bibitem{Aglinet2000}
Agilnet.
\newblock {Understanding and Improving Network Analyzer Dynamic Range}.
\newblock Technical report, Agilnet, 2000.

\bibitem{GORDON1988}
Clair~C Gordon, Thomas Chirchill, Charles~E Clauser, Bruce BradtMiller, John~T
  McConville, Ilse Tebbetts, and Robert~A Walker.
\newblock {Anthropometric Survey Of U.S. Army Personnel, Summary Statistics,
  Interim Report}.
\newblock {\em NATICK/TR-89/027}, 1988.

\bibitem{Gabriel1996}
S.~Gabriel, R.~W. Lau, and C.~Gabriel.
\newblock {The dielectric properties of biological tissues: III. Parametric
  models for the dielectric spectrum of tissues}.
\newblock {\em Physics in Medicine and Biology}, 1996.

\bibitem{Gabriel1996a}
C.~Gabriel, S.~Gabriel, and E.~Corthout.
\newblock {The dielectric properties of biological tissues: I. Literature
  survey}.
\newblock {\em Physics in Medicine and Biology}, 1996.

\bibitem{Gabriel1996b}
S.~Gabriel, R.~W. Lau, and C.~Gabriel.
\newblock {The dielectric properties of biological tissues: II. Measurements in
  the frequency range 10 Hz to 20 GHz}.
\newblock {\em Physics in Medicine and Biology}, 1996.

\bibitem{Chollet2015}
Fran{\c{c}}ois Chollet.
\newblock {Keras: Deep Learning library for Theano and TensorFlow}, 2015.

\bibitem{Kingma2015}
Diederik~P. Kingma and Jimmy~Lei Ba.
\newblock {Adam: A method for stochastic optimization}.
\newblock In {\em 3rd International Conference on Learning Representations,
  ICLR 2015 - Conference Track Proceedings}, 2015.

\bibitem{IanGoodfellowYoshuaBengio2015}
Aaron~Courville {Ian Goodfellow, Yoshua Bengio}.
\newblock {Deep Learning Book}.
\newblock {\em Deep Learning}, 2015.

\bibitem{Salzberg1997}
Steven~L. Salzberg.
\newblock {On comparing classifiers: Pitfalls to avoid and a recommended
  approach}.
\newblock {\em Data Mining and Knowledge Discovery}, 1997.

\bibitem{Abubakar1997}
Aria Abubakar, Tarek~M Habashy, Richard~F Bloemenkamp, Den Berg,
  Solving~Electromagnetic Inverse, Cross-correlated~Contrast Source, and
  Inversion Method.
\newblock {Related content A contrast source inversion method}.
\newblock {\em Inverse Problems}, 13:1607--1620, 1997.

\bibitem{Mojabi2012}
Puyan Mojabi, Majid Ostadrahimi, Lotfollah Shafai, and Joe LoVetri.
\newblock {Microwave tomography techniques and algorithms: A review}.
\newblock In {\em 2012 15th International Symposium on Antenna Technology and
  Applied Electromagnetics}, 2012.

\bibitem{Semenov2009}
Serguei Semenov.
\newblock {Microwave tomography: Review of the progress towards clinical
  applications}.
\newblock {\em Philosophical Transactions of the Royal Society A: Mathematical,
  Physical and Engineering Sciences}, 367(1900):3021--3042, 2009.

\bibitem{Semenov2017}
S.~Semenov, Yu~Portnov, A.~Semenov, A.~Korotkevich, and A.~Kokov.
\newblock {Neuroimaging patterns of cerebral hyperperfusion}.
\newblock {\em Journal of Physics: Conference Series}, 886(1), 2017.

\bibitem{Powers2019}
{Guidelines for the early management of patients with acute ischemic stroke:
  2019 update to the 2018 guidelines for the early management of acute ischemic
  stroke a guideline for healthcare professionals from the American Heart
  Association/American Stroke Association}, 2019.

\end{thebibliography}






\end{document}